\documentstyle[floats,aps]{revtex}
\begin{document}
\input epsf
\draft

\twocolumn[\hsize\textwidth\columnwidth\hsize\csname  
@twocolumnfalse\endcsname
\title{ e-h Coherence and Charging Effects in Ultrasmall Metallic Grains}

\author{S. Drewes and S.R. Renn}
\address{
Department of Physics, University of California at San Diego,
La Jolla, CA 92093}
\author{F. Guinea}
\address{Instituto de Ciencia de Materiales, 
Consejo Superior de Investigaciones Cient{\'\i}ficas,
Cantoblanco, E-28049 Madrid, Spain}

\date{\today}

\maketitle

\begin{abstract}
We consider a model for electron tunneling between a pair of ultrasmall
metallic grains. Under appropriate circumstances, non-equilibrium final state
effects can strongly enhance tunneling and produce electron-hole coherence
between the grains. The model displays a quantum phase transition between a
Coulomb blockaded state to a coherent state exhibiting subohmic tunneling
conductance.  The critical state of the junction
exhibits a temperature independent resistance of order $h/e^2$.  Finally
we discuss the possible relevance to granular materials and quantum dots.
In particular, similarities between the quantum transition in our model and
the metal-insulator transition in granular wires observed by Herzog
{\it et al.\/} are described in detail.
\end{abstract}

\vskip2pc]

\narrowtext

Recent work by Herzog {\it et al.\/} \cite{Herzog} have found dramatic
evidence for an unusual metal insulator(MI) transition in granular wires
fabricated by {\it in situ\/} deposition through a metallic stencil onto a
GaAs substrate.  The transition involves an abrupt multi-order change in wire
resistance as a function of the amount of deposited metal. The transition
occurs in a variety of materials (including Sn, Pb, Au, Ag, and
Pb$_{0.85}$Bi$_{0.15}$).  Although the transition may be observed in wires
as wide as $7000$\AA, the resistance gap decreases with increasing wire width
and is absent in two-dimensional films.

In this paper, we wish to consider the possibility of modeling the above
behavior in terms of  a pair of metallic grains in close proximity.
Our model will include two ingredients.
The first is electrostatic charging effects.  This should be important  
in the Herzog {\it et al.\/} experiments since the charging energy $E_Q$
is estimated\cite{estimate} to be large (100K).  The second ingredient is a
non-equilibrium final state effect in which electrostatic fields between
the two particles are suddenly switched on during the tunneling process.
This effect is an exciton effect in which the tunneled electron
is attracted to the positively charged counter-electrode. 
We will show that the competition between the
exciton effect and Coulomb blockade gives rise to a MI transition  between
a phase exhibiting sub-Ohmic $I(V)$ characteristics to a phase exhibiting 
a Coulomb blockade. The critical state separating these two phases exhibits
a temperature independent conductance.  Finally, we will discuss the
the similarities between these results and those of Herzog {\it et al.\/}
\cite{Herzog}.

\begin{figure}[t]
\centering
\leavevmode
\epsfxsize=9cm
\epsfbox[0 0 493 619] {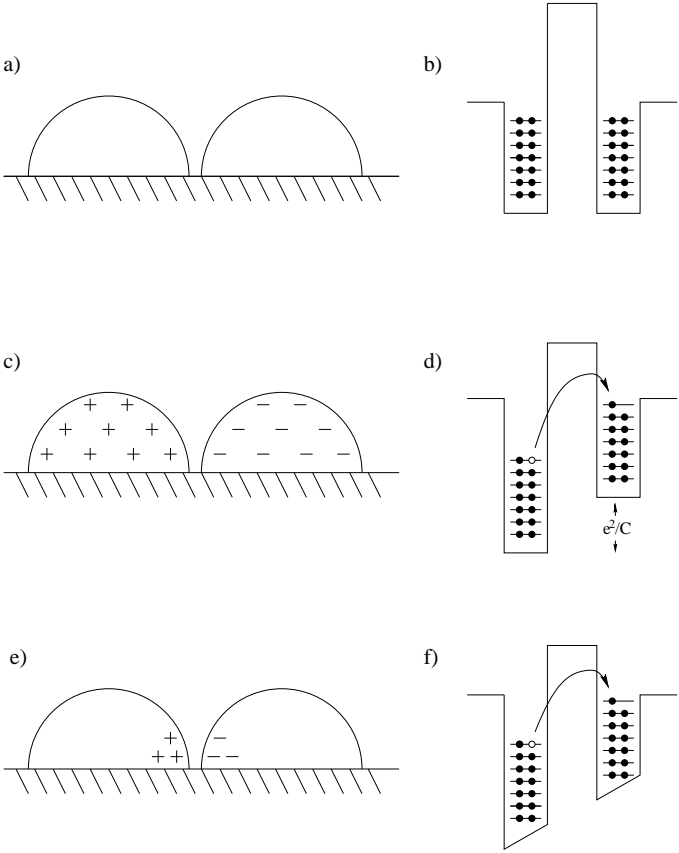}
\caption[]
{\label{tunnel}
(a) An illustration of two nearby metallic grains on an
insulating substrate. The two grains form an ultrasmall  tunnel junction.
Although the grains may be part of a composite granular material, coupling to
other grains in the host material will be ignored. (b) Energy levels and
confinement potential associated with (a). (c) Metallic grains after
tunneling event.  An unrealistic charge distribution 
is obtained when  $V^{\alpha}_{k k'}$ is set to zero.
(d) The energy levels and confinement potentials associated with (d).
(e) Metallic grains after tunneling event.  Charges localized at the tunneling
site give rise to long range electrostatic interactions which act as
a suddenly switched-on potential $V^{L}_{k k'}=V^R_{k k'}$.
(f) The non-equilibrium energy levels and confinement potential
associated with (e).}
\end{figure}

We begin our discussion by considering a pair of  identical
metallic grains on an insulating substrate (see Fig. \ref{tunnel}).
The grains are in close proximity and form an ultrasmall tunnel junction with 
intergrain capacitance $C<10^{-15}$F (For the present argument, we will neglect
the intragrain capacitance).  The two grains may be part of a granular host.
However, tunneling to other grains in the host material will be ignored.
Now consider a tunneling process in which an electron tunnels from grain \#1
to grain \#2 (see Fig. 1c-1d).  In the classical Coulomb blockade picture 
a tunneled electron causes all energy levels in the grain \#2 to up-shift by
$e^2/2C$ and all energies in grain \#1 to down shift by $e^2/2C$.  In the
diagram, we have represented the electrostatic potentials and surface
confinement potential as a square well whose shape is unaffected by the
tunneling process.  Notice that the classical Coulomb blockade picture does
not properly describe the non-equilibrium effect associated with suddenly
switching on the electrostatic attraction between the two grains (see Fig.
1e-1f).  Such effects can give rise to shake-up (orthogonality catastrophe)
effects which can seriously effect tunneling rates \cite{Guinea,Ueda}
especially in small particles.  To see how important these effects might be,
consider the classic problem of the x-ray absorption edge \cite{mahan,Noziers}. 
In that problem, the absorption intensity, near threshold, could vanish at
threshold due to the orthogonality catastrophe\cite{Noziers} or could exhibit
a power law divergence known as the exciton effect\cite{mahan}.
{\it A priori\/} one might expect that a similar effect could cause the
differential conductance of a tunnel junction to vanish or diverge.

To examine these effects in detail, we consider a model
\cite{BenJacob,Guinea,Strohm,Ueda} which describes tunneling between the two
grains,
\begin{equation}
H=H_L+H_R+H_T+\frac{Q^2}{2C}
\label{pacoH}
\end{equation}
where $H_T=\sum_{kp} [T_{kp}e^{i\phi}c^+_kc_p+{\rm h.c.}]$ and where $\phi$
is defined in terms of the voltage difference $V_L-V_R$ across the grains
via the relation $\dot{\phi}=e(V_L-V_R)/\hbar$. Finally, the $H_L$ and $H_R$
are given by
\begin{equation}
H_{\alpha}=\sum_{k} \epsilon^0_{k,\alpha} c^{+}_{k \alpha}c_{k \alpha}
+Q\sum_{k k'} V^{\alpha}_{k k'} c^+_{k \alpha} c_{k' \alpha}
\end{equation}
where $V^{\alpha}_{k k'}$ represents the sudden change of surface and
electrostatic potentials on grain $\alpha=L,R$ which occurs during the
tunneling process. If $V^{\alpha}_{k k'}=0$, we recover the standard model
\cite{BenJacob} (see Figs. 1c-1d and 2a) which describes the effects of
particle hole excitations induced by tunneling processes in
ultrasmall tunnel junctions. $V^{L}_{k k'}=V^R_{k k'} \ne 0$  would be chosen
to include shake-up and other final state effects in symmetric tunnel junctions.

Now consider the zero temperature tunnel conductance to leading
order in $T_{k p}=T$.  One may calculate the tunneling current using
$I=-2e {\rm Im}\{X_{ret}(-eV)\} $ where $X_{ret}=|T|^2 \int_0^{\infty} d \tau 
G^e_R(\tau)G^h_L(\tau) \exp{i \omega\tau}$ is the retarded response function
associated with the non-equilibrium production of the R-electron and L-hole.
Now the non-equilibrium electron and hole propagators associated with the
suddenly switched on $V^{\alpha}_{k k'}$ are given by 
$G^{h,e}_R(t) \propto N_R t^{-(1\pm\delta_R/\pi)^2} \exp{-iE^R_f t}$
where $N_{\alpha}$ is the DOS at the Fermi level in the $\alpha=L,R$
electrode and $\delta_{\alpha}$ is
a phase shift associated with the scattering of electrons off the potential
$V^{\alpha}_{k k'}$.  In the above result, $+$ is used for hole and $-$ for
electrons in the right electrode.  Similarly $G^{h,e}_L(t) \propto N_L
t^{(1 \mp \delta_L/\pi)^2} \exp -iE^L_ft$ for the left electrode.  Because
of intergrain charging energy, the electrons tunnel to a non-equilibrium
state characterized by Fermi levels shifted such that $E^R_f-E^L_f=e^2/2C$.
Combining these results one obtains $\frac{dI}{dV} \propto
\frac{1}{R_T}\left(\frac{e^2/C}{eV-e^2/2C}\right)^{\epsilon}$
where $\epsilon=2(\delta_R/\pi +\delta_L/\pi)-(\delta_R/\pi)^2 
-(\delta_L/\pi)^2$ and $R^{-1}_T= 4\pi e^2 N_LN_R|T|^2$
For small positive $\delta_{\alpha}$, there is a competition between
excitonic effects associated with $2(\delta_R/\pi +\delta_L/\pi)$ and
orthogonality effects associated with $-(\delta_R/\pi)^2 -(\delta_L/\pi)^2$.
Depending on which terms dominate, one can obtain (Fig. 2b) a divergent
differential conductance \cite{oops} at threshold $V_T=e^2/2C$ or (Fig. 2c)
a vanishing conductance at threshold. 

\begin{figure}[t]
\centering
\leavevmode
\epsfxsize=9cm
\epsfbox[18 18 552 482] {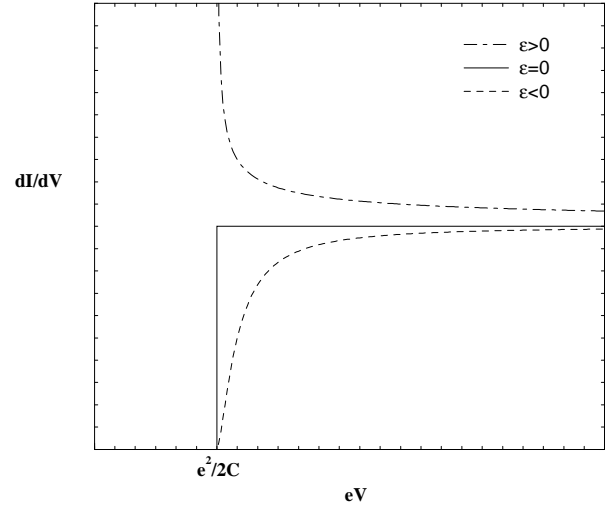}
\caption[]
{\label{text}
Differential conductance for $\epsilon<0$ (dot-dashed), $\epsilon=0$
(solid) and $\epsilon>0$ (dashed). $\epsilon>0$ curve assumes exciton effect
dominates the orthogonality catastrophe.}
\end{figure}

In general the  form of the final state interactions $V^{\alpha}_{k k'}$ is
not important to the following discussion and is difficult to calculate.
However there are several observations which can be made:
First we observe that one can estimate\cite{MGB,MAHAN} $|\delta_{L,R}|
\sim \pi/k$ where $k$ is the number of transverse channels available for
intergrain tunneling.  Next, we observed  that $\delta_R$ will be positive if
$V^{\alpha}_{k k'}$ is a potential which tends to keep the tunneled electron
in the electrode near the tunneling site {\it i.e.\/} if the electron is
attracted to the positively charged electrode. Similarly $\delta_L$ will be
positive if the hole is attracted to the negatively charged electrode. 
Hence the electrostatic  interaction between a pair of grains with a small
number of accessible tunneling channels is expected to give a positive
$\delta_{\alpha}$ large enough to make exciton effects observable.

To some readers it may be surprising that the repulsive Coulomb interactions
would enhance tunneling between the grains. The behavior is not unusual and
can be found in several simple models. 
For instance, consider a pair of semi-infinite 1-D spinless chains 
described by the Hamiltonian
$H=[t'c^+_{L 0}c_{R 0}+h.c.]+\sum_{i=0} [tc^+_{L i+1}c_{Li}+h.c. ]+
U(c^+_{L}(0)c_L(0)-1/2)(c^+_{R}(0)c_R(0)-1/2)$
This model is equivalent to a 1-D Anderson impurity model.  The
interchain tunneling is associated with a transverse magnetic field acting
on the impurity.  For  $U>0$ the magnetic susceptibility and transverse
magnetization $\langle c^+_{L}c_R\rangle$ will be rapidly enhanced with
increasing $U$.  Hence, the differential tunneling conductance 
will be significantly increased by a large positive $U$. 

The above discussion of the exciton effect has been performed to leading 
order in $|T|^2$.  We will now go beyond leading order in $|T|^2$ and show
that if $|T|$ large and $\epsilon>0$, that the Coulomb blockade is destroyed.
To do this, we integrate out the particle-hole excitations within the grains. 
This gives an effective action in imaginary time of the form
\begin{eqnarray*}
S&=&\int^{\beta \hbar}_0 d\tau \ \frac{C}{2e^2}\dot{\phi}^2+\\
& &  \tau_Q^{-\epsilon}\int^{\beta \hbar}_0 d\tau 
d \tau' \alpha(\tau-\tau')[1-\cos(\phi(\tau)-\phi(\tau'))]
\end{eqnarray*}
where $\alpha(\tau)=\alpha_0 \left(\frac{\pi k_B T}{\sin(\pi k_B T \tau)}
\right)^{2-\epsilon}$ and $\alpha_0=\hbar/(2\pi e^2 R_T)$.
This is a one-dimensional XY model with long range interactions.  This model
had been studied within the framework of the renormalization group by
Kosterlitz\cite{Kosterlitz} who found an order-disorder transition at
$\alpha_0=\alpha_{\rm c}=2/(\epsilon \pi^2)$ for $\epsilon \ge 0$. The model is
disordered for $\epsilon \le 0$ although the absence of an ordered phase
when $\epsilon=0$ was a source of controversy.\cite{Scalia,Simanek,Zwerger}.

In order to understand the nature of the two phases,
we calculated the conductance of the model using the Kubo formula
$G(\omega)=\langle |I_t(\omega)|^2\rangle/\omega$ where $I_t(\omega)$ is
the tunneling current.  To leading order in an expansion in powers
of $1/\alpha$, a spin-wave calcualtion reveals that
\begin{equation}
G=\frac{2\sqrt{\pi}\alpha_0}{R_Q}\frac{\Gamma((1+\epsilon)/2)}
{\Gamma(1+\epsilon/2)}\left(\frac{E_Q}{\pi k_B T}\right)^{\epsilon}
\label{Gdiverg}\end{equation}
where $E_Q=e^2/2C$ and $R_Q=\hbar/e^2=4.11k\Omega$.  We see that $G$ diverges
at $T\rightarrow0$. Hence one identifies the
ordered phase as subohmic. One can also calculate the conductance to leading
order in $\alpha_0$. In this case one finds
$G\sim \frac{\alpha^2_0}{R_Q} \left(\frac{\pi k_B T}{E_Q}
\right)^{2(1-\epsilon)}$ which vanishes as $T\rightarrow 0$ indicating that
the disordered phase is insulating.

\begin{figure}[h]
\centering
\leavevmode
\epsfxsize=9cm
\epsfbox[18 18 552 482] {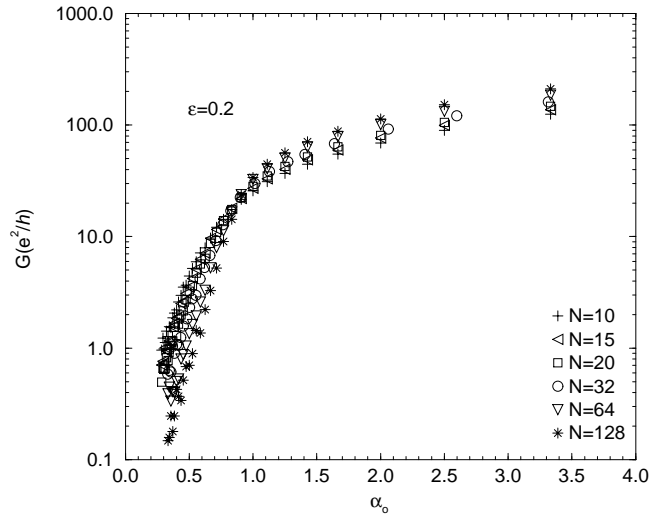}
\caption[]
{\label{cond.min.2} Conductance obtained from a MC simulation.  Notice,
$G(\alpha_0)$ curves intersect at $G_{\rm c} \approx 11e^2/h$. This suggests
that the conductance of the critical state is a universal, temperature
independent constant.}
\end{figure}

To understand the transition in greater detail, we have evaluated the DC
conductance using a Monte Carlo simulation\cite{timeslice}.
Using\cite{Simanek}
\begin{equation}
G=\frac{2\pi \alpha_0}{\hbar \beta R_Q}\int_0^{\hbar \beta}
\gamma_{\epsilon}(\tau)\langle \cos(\phi(\tau)-\phi(0))\rangle \ d\tau
\label{GDC}\end{equation}
where  $\gamma_{\epsilon}(\tau)=[\pi (k_B T/E_Q)/
\sin(\pi k_B T \tau \hbar)]^{-\epsilon}$, we evaluate $G$ using
$\langle \cos(\phi(\tau)-\phi(\tau'))\rangle$
obtained from a series of simulations including $10^5$ cycle runs for the
10, 20, and 32 timeslice systems and $2\times 10^5$ cycle runs for the 64 and
128 timeslice systems. The results for $\epsilon=0.2$ are presented in
Fig. \ref{cond.min.2}  The transition between the 
sub-Ohmic (high $\alpha_0$) phase to the insulating (low  $\alpha_0$)
phase is evident. Interestingly one observed the conductance curves cross
at a single point. This point identifies a transition at a critical value of
$\alpha_{\rm c}=0.9$ which compares well to $\alpha_{\rm c}=1.01$ obtained by
Kosterlitz RG treatment.

The fact that curves intersect at all indicates that $\alpha_{\rm c}$ separates
metallic  from insulating behavior.  However, the observation that all line
cross at $G_{\rm c}\approx 11\frac{e^2}{h}\approx 1/(2.3k\Omega)$ seems to
indicate that the critical state has a finite temperature independent
conductance. The existence of a critical state with a finite resistance can
be understood as follows.  In general, finite size scaling theory
implies that the critical states exhibits correlations of the form
$\langle\exp i\phi(\tau) \exp -i \phi(0)\rangle =\left( \frac{\tau_Q}{\tau}
\right)^{d-2+\eta}F(\hbar \beta/\tau)$
where $\tau_Q=\hbar/E_Q$ is the width of the time slices, $d=1$ is the
space-time dimensionality, and where $F(x)$ is a universal scaling function
which is finite as $x\rightarrow 0$. According to Fisher, Ma and Nickel
\cite{Fisher}, $\eta=1+\epsilon$ is exact for our model.  It follows that
\begin{equation}
G_{\rm c}=b(\epsilon)\frac{e^2}{h}
\label{Gc}\end{equation}
at criticality. Since the DC conductance depends only on the
$\omega\rightarrow 0$ limit of the model, the above result is a universal but
$\epsilon$ dependent result. Typically simulations for other values of
$\epsilon$ reveal that $1/G_{\rm c} \sim 10^5\Omega$ except in the 
$\epsilon \rightarrow 0$ limit where $G_{\rm c}\rightarrow \infty$.
It should be mentioned that the Herzog data reveals that the highest
resistance metallic state in Au (width: $400\AA$), Pb$_{0.85}$Bi$_{0.15}$
(width: $575\AA$ and $850\AA$), and Sn (width: $550\AA$) wires
have critical resistances of $2k\Omega$, $2k\Omega$, $4k\Omega$, and 
$4k\Omega$, respectively. Such values of $G_C^{-1}$ are  consistent with
eqn. \ref{Gc}.

\begin{figure}[!h]
\centering
\leavevmode
\epsfxsize=9cm
\epsfysize=9cm
\epsfbox[0 0 548 467] {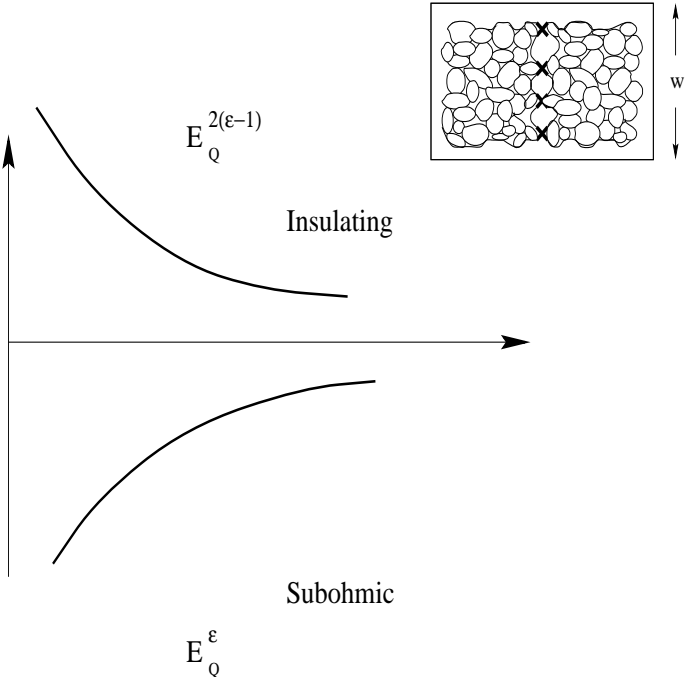}
\caption[]
{\label{grain} Modeling a crack as a series of parallel tunnel junctions
(see inset).
Observe that the charging energy decreases with increasing wire width. The
decreasing $E_Q$ causes the resistance gap to close and the resistance of
non-critical subohmic wires to increase.}
\end{figure}

\noindent {\it Discussion --}\/ We wish to consider the relevance of
our tunnel junction to the MI transition observed by Herzog {\it et al.\/}
\cite{Herzog}.  Herzog's MI transition occurs in many different materials
and exhibits a resistance gap which depends on the wire width, 
$500\AA-6000\AA$. This suggests that the transition does not involve some 
microscopic phenomenon
in the interior/surface of  the grains but, instead, involves some low energy
collective behavior which requires  two or more grains. In addition, it
appears that the  MI transition is
not a  percolation transition in which the drop in the
resistance is determined by purely geometrical considerations such as number
of parallel connections spanning a backbone cluster. This is clear since
percolation transition  can occur in films
but not in  wires. At this point, one might model
the wire as a disordered network of tunnel junctions
where Coulomb blockade effects could produce an insulating phase. This is
reasonable since charging energies are estimated\cite{estimate} to be
$\sim 100K$.  Assuming the tunnel junction conductances to be broadly
distributed, the wire resistance will be dominated by
the tunnel junctions with highest resistance \cite{Stauffer}.

Now consider the dependence of the wire resistance on the wire width.  Let
$l_{\phi}$ be the phase coherence length {\it i.e.\/} the length scale that
correlations $\langle e^{i\phi(\vec{r})}e^{-i\phi(0)}\rangle$ die off in the
disordered phase.  Then for wire widths $w\ll l_{\phi}$, the low frequency
behavior of the obstruction will be described by the tunnel junction model with
a charging energy which  decreases with increasing 
wire width. For instance,  in a narrow wire ($w\ll l_{\phi}$)
phase difference across multiple tunneling sites spanning a crack or weak-link
will be equal. (See fig. 4.) Hence, the  capacitances comprising the weak-link
add in parallel.  This is a useful observation since the decreasing
charging energy associated with increasing wire width
will cause the resistance of the non-critical subohmic states to increase
like $(T/E_Q)^{\epsilon}$ (eq. \ref{Gdiverg}) similar to the behavior observed
by Herzog {\it et al.\/}. The increase of the metallic wire resistance with
wire width is a unique phenomenon which is difficult to obtain from
alternative models.  The model also predicts that the resistance of
insulating wires will {\it decrease} with increasing wire width.
Consequently the resistance gap will decrease until $w\sim l_{\phi}$
where the crossover to two dimensional transport will occur.

At this point, we should mention that  our model makes several 
predictions which have not yet been tested. First, from the
Kosterlitz RNG treatment demonstrates the existence of a characteristic
energy scale, $\Delta =E_Q(1-\alpha_0/\alpha_{\rm c})^{1/\epsilon}$. 
This behavior should be detectable by examining the differential conductance
which obeys a scaling ansatz of the form 
$dI/dV =\frac{e^2}{h}(eV/E_Q)^{\delta}F(eV/\Delta)$ where $\delta=0$
in order that scaling ansatz is consistent with eq. \ref{Gc}.
The second  prediction is that the transition to the phase
coherent state in arrays of coupled tunnel junctions  will be accompanied by
a divergence in $\l_{\phi}$.
This unique behavior should be readily observable in magnetoresistance 
measurements performed on insulating granular wires.

We should also mention that one should also be able to
search for the subohmic to insulator transition
in  double quantum dot systems of the sort considered by Waugh {\it et al.\/}
\cite{Waugh,Waughcomment}  The double-dot systems
have several useful features including (1.)a small number of tunneling channels
which implies the large phase shifts\cite{MGB} required for an
exciton effect, (2.) small intergrain capacitances\cite{Waugh}, and 
(3.) precise control of the tunneling barrier between dots.
Unfortunately, the Waugh  experiment itself could not distinguish between a
cross-over and a phase  transition. However, this is
not an inherent limitation of the experimental method. 
So an  attempt to search for this quantum phase transition 
in the double dot should be feasible and would certainly  be most welcome.

\noindent {Acknowledgements} The authors would like to acknowledge support from
NSF Grant No. DMR 91-13631 (S.R.),  the Hellman foundation(S.R.), 
the Alfred P. Sloan Foundation (S.R.). We would like to acknowledge
useful conversations with D. Arovas, R.C. Dynes, M. Ueda, F. Waugh, and 
P. Xiong.


\begin{references}
\bibitem{oops}The possibility of a divergent conductance is in contrast to
a previous result by one of us\cite{Guinea} which 
concluded that the exciton effect
vanished for symmetric tunnel junctions. That result  was based on 
the assumption that the electrostatic potentials acting on the electron and
hole were identical. This is not the case since the electron and hole
propagate in different grains  and experience electrostatic potentials of
opposite polarity (see Fig. 1e-1f).
\bibitem{timeslice} $S_{eff}$  without
the charging term was simulated on a 1-D lattice of spacing $\tau_0=\hbar/E_Q$.
\end{references}
\end{document}